\documentclass{iaus}

\usepackage{graphicx}

\title[The PSU-TCfA Search for Planets Around Evolved Stars] 
{The PSU/TCfA Search for Planets around Evolved Stars. Stellar parameters and activity indicators of targets.}

\author[A. Niedzielski, G. Nowak, \& P. Zieli{\'n}ski]   
{Andrzej Niedzielski$^{1, 2}$, Grzegorz Nowak$^1$, \and Pawe{\l} Zieli{\'n}ski$^1$}

\affiliation{$^1$Toru{\'n} Centre for Astronomy, Nicolaus Copernicus University, \\ ul. Gagarina 11, 87-100 Toru{\'n}, Poland, \\ email: {\tt aniedzi, grzenow, pawziel @astri.uni.torun.pl} \\[\affilskip]

$^2$Department of Astronomy and Astrophysics Pennsylvania State University, \\ 525 Davey Laboratory, University Park, PA 16802 

}

\pubyear{2008}

\volume{249}  

\pagerange{100--104}

\jname{Exoplanets: Detection, Formation and Dynamics}

\editors{A.C. Editor, B.D. Editor \& C.E. Editor, eds.}

\begin{document}

\maketitle

\begin{abstract}

The main objective of the Penn State/Toru\'n Centre for Astronomy Search for Planets around Evolved Stars is  the detection  of planetary systems around massive, evolved stars. We are also interested in  the evolution of these systems on stellar evolution timescales. In this paper we present our approach to determine the basic physical parameters of our targets GK-giants.  We also discuss the stellar activity indicators used in our survey: line bisector and curvature, and H$\alpha$ variability.



\end{abstract}

\firstsection 


\section{Introduction}

Proper interpretation of the results from precision RV studies of GK-giants requires a detailed knowledge of their physical parameters.  Effective temperatures and gravitational accelerations are needed to obtain luminosities, and, with the additional knowledge of metallicities, estimates of stellar masses and ages can be derived by means of the isochrone fitting. Together with estimates of stellar radii and rotation periods, these data allow one to address the influence of stellar surface inhomogeneities (spots) on the observed RV variations.

All alternative sources of RV variations in GK-giants have to be ruled out before substellar companion interpretation becomes acceptable. Unfortunately, the long period variations, if present, cannot usually be studied using  data other than the existing photometry (usually of moderate quality), or the data collected from the RV survey. Therefore, the detailed activity discussion is usually based on the indicators defined on the basis of the same spectra as those used for the RV measurements. In this paper, we illustrate our methodology of the determination of stellar parameters and activity analysis based on the case of the K-giant PSU-TCfA 18, a potential planet hosting star.

\section{Observations}

Observations were made with the Hobby-Eberly Telescope (HET) and the High Resolution Spectrograph (HRS). The HRS was used in the R=60,000 resolution mode with a gas cell ($I_2$) inserted into the optical path, and it was fed with a 2 arcsec fiber. Typically, the signal-to-noise ratio per resolution element (at 594 nm) was $\sim$200 for the stellar spectra taken with the gas cell, and $\geq$250 for the templates.

\section{Basic stellar parameters}

The atmospheric parameters of the program stars were obtained with the spectroscopic method (\cite[Takeda \etal\ 2005a, b]{Takeda2005a,Takeda2005b}), which is based on analysis of Fe~I and Fe~II lines and relies on conditions resulting from the assumption of the LTE. Typically, over 200 FeI and about 25 FeII lines were measured for every star.

We tested a reliability of our determinations  with the \cite[Takeda~{\it et~al.} (2005a)]{Takeda2005a} TGVIT code by applying it to 8 stars, for which the parameters have been published by \cite[Butler \etal\ (2006)]{Butler2006}. A comparison of the results shows that $T_{eff}$ values agree to within 49 K, and that the same is true for $\log~\emph{g}$ to within 0.11 dex and $[Fe/H]$ to within 0.11 dex, respectively.

 Stellar masses were derived by comparing  the positions of stars in the HR diagram with the theoretical evolutionary tracks of \cite[Girardi \etal\ (2000)]{Girardi2000} and \cite[Salasnich \etal\ (2000)]{Salasnich2000} for a given metallicity. For stars for which the parallax determinations are precise enough, the metallicity may introduce a significant uncertainty in mass because of the choice of an evolutionary track. We assume that for an average red giant with a known parallax, the mass may be estimated to within $0.3 M_{\odot}$. We also note that for stars in the red giant clump, which are  in the fast evolution phase with mass-loss, the derived masses are probably the upper limits.

Stellar radii were determined using the calibration given in \cite[Alonso \etal\ (2000)]{Alonso2000}.  Stellar ages are usually estimated with the aid of the theoretical stellar isochrones.

 In the case of the PSU-TCfA~18 star,  we have measured equivalent widths of up to 195 Fe~I and 11 Fe~II lines for further analysis.  For this star, the basic physical parameters are $T_{eff} = 4246 K$, $\log~\emph{g} = 2.43$, $v_{t} = 1.52 km s^{-1}$, $[Fe/H] = 0.11$, $M = 5.5 M_{\odot}$, and $R = 27.4 R_{\odot}$. Intrinsic uncertainties of our determinations are $\sigma T_{eff} = 29 K$, $\sigma \log~\emph{g} = 0.09$, $\sigma v_{t} = 0.13 km s^{-1}$ and $\sigma [Fe/H] = 0.06$.

\section{\boldmath $V \sin i$ measurements}

Rotation periods represent a parameter of particular importance in searches for planetary companions to red giants. As the rotation periods of these stars are very similar to expected orbital periods, their knowledge is critical for an unambiguous interpretation of observations. Any correlation of stellar activity indicators variations with the rotation period make a substellar companion hypothesis unlikely. To estimate rotation periods from our spectra we have used the cross-correlation technique, as described in \cite[Benz \& Mayor (1984)]{BandM1984}.

We have computed  the CCFs by cross correlating the high S/N blue spectra with a numerical mask. To measure $V \sin i$ we have worked out  a $V \sin i$ calibration for the HET/HRS. To determine the $\sigma_{0}$ vs. $(B-V)$ relationship,  we have used 16 slow rotators with accurately known projected rotational velocities, preferably from  \cite[Gray (1989)]{Gray1989}, \cite[Fekel (1997)]{Fekel1997} and \cite[de Medeiros \& Mayor (1999)]{deMandM1999}. For  these stars we have determined $\sigma_{0}$ using the formula from \cite[Benz \& Mayor (1984)]{BandM1984} ($V \sin i = A \sqrt{\sigma_{obs}^{2} - \sigma_{0}^{2}}$ ) and assuming the constant $A = 1.9$ following the \cite[Queloz \etal\ (1998)]{Queloz1998} and \cite[Melo \etal\ (2001)]{Melo2001}. We have carried out a least-squares fit to the  data of the analytical function $\sigma_{0} = a_{2} (B-V)^{2} + a_{1} (B-V) + a_{0}$, which  yields the following calibration: $\sigma_{0} = 15.592(B-V)^{2} - 26.753(B-V) + 14.559$.

 Using this calibration we have obtained $V \sin i = (3 \pm 1) km s^{-1}$ for our star. Adopting the radius for this star as determined above, we have estimated its rotational period to be 220 - 950 days. The large uncertainty in the  rotation period is caused by  uncertainties related to the determination of the radius and $V \sin i$.

\section{Stellar activity indicators}

One of  possible sources of the observed RV variations in GK-giants is due to their pulsations. Therefore,  photometric data that span long periods of time are needed for the interpretation of the results of our survey. Because we do not conduct our own parallel photometric observations,  we must rely on the existing photometric databases like {\it Hipparcos} or {\it NSVS} (\cite[Wo{\'z}niak et al. 2004]{2004AJ....127.2436W}). These moderate quality data  provide time-series, which are long enough to be useful in searches for long-term periodicities. However, these measurements  were usually performed many years prior to our RV survey. In the particular case of the PSU-TCFA 18 star, no detectable variability is present in the existing photometric data.

\subsection{Line bisectors}

The basic tool to study the origin of RV variations derived from the stellar spectra is the analysis of the shapes of spectral lines via line bisectors (\cite[Gray 1983]{Gray1983}).

We have computed line bisectors for 5 strong, unblended spectral features of a moderate intensity, which were located close to the center of echelle orders: Cr I 663.003 nm, Ni I 664.638 nm, Ca I 671.77 nm, Fe I 675.02 nm, and Ni I 676.784 nm. All these lines show well defined bisectors.

The changes in the spectral line bisectors were quantified using the bisector velocity span ($BVS$) parameter, which is simply the velocity difference between the upper and the lower points of the line bisector ($BVS = v_3 - v_2$), and the bisector curvature ($BC$), which is the difference of the velocity span of the upper half of the bisector and its lower half ($BC = (v_3 - v_2) - (v_2 - v_1)$). It is important to examine both $BVS$ and $BC$, because it is possible for a star to show variations in one of these parameters only. In choosing the span points, it is important to avoid the wings and cores of the spectral line, where the errors in the bisector measurements are large. For our span measurements we chose $v_1=0.29$, $v_2=0.57$, and $v_3=0.79$ in terms of the line depth at the line core. Using the bisector measurements of all 5 spectral lines we have computed the average velocity span and curvature after subtracting the mean value for each spectral line.

\begin{figure}[h]

\begin{center}

\includegraphics[width=1.9in]{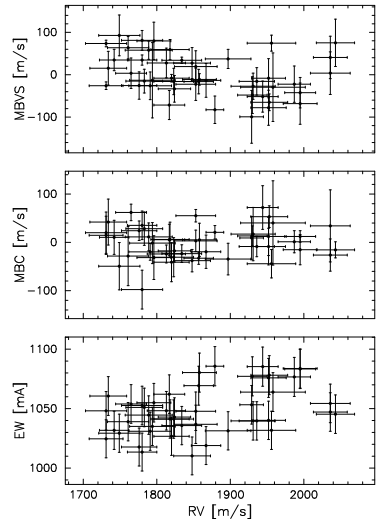} 

\caption{MBVS, MBC and EW of the $H \alpha$ for PSU-TCfA 18 as a function of radial velocity.}

\label{fig1}

\end{center}

\end{figure}

In Figure\,\ref{fig1}, we present the mean bisector velocity span ($MBVS$)and the mean bisector curvature ($MBC$) for our star, as a function of RV. Uncertainties in the derived values of  $MBVS$ and $MBC$ were estimated as standard deviations of the mean. The correlation coefficients were found to be $r=-0.32 \pm 0.10$ for $MBVS$ and $r=0.07 \pm 0.02$ for $MBC$. It is clear that they are not correlated with radial velocities which supports the planetary mass companion hypothesis. The available RV measurements are not uniformly distributed over the estimated period, which is visible in Fig. 1 as a scatter varying with RV. More observations are needed to confirm the apparent lack of correlation.

\subsection{ $H \alpha$ variations}

Since our spectra start at 407.6 nm we cannot investigate the variation of the Ca II K emission line (393.4 nm). Also the infrared CaII triplet lines 849.8-854.2 are outside the range of our spectra. Therefoer, we use H$\alpha$ line (656.28 nm) as  a chromospheric activity indicator. The EW measurements of the $H \alpha$  line can be made in our spectra  with a typical precision of a few percent. In the case of our star, the measurements give a mean value of $1049 \pm 20 m\AA$. The rms value of $20 m\AA$ corresponds to 2 \% variation in the EW. In Figure \,\ref{fig1} we present EW measurements for $H \alpha$ as a function of RV. The correlation coefficient of  $r = 0.44 \pm 0.10$ shows marginal relationship (probably resulting from the non-uniform RV coverage) which again supports the planetary hypothesis.

\section{Conclusions}

A detailed knowledge of stellar parametres of red giants is very important for interpretation of their RV variations. To rule out stellar activity as the source of such variations, one needs  precise rotation periods and several other indicators to be measured at many epochs.

\section*{Acknowledgements}

AN \& GN acknowledge the financial support from the MNiSW through grant 1P03D 007 30. GN is a recipient of a graduate stipend of the Chairman of the Polish Academy of Sciences. PZ was supported by MNiSW grant SPB 104\/E-337\/6. The Hobby-Eberly Telescope (HET) is a joint project of the University of Texas at Austin, the Pennsylvania State University, Stanford University, Ludwig-Maximilians-Universit\"at M\"unchen, and Georg-August-Universit\"at G\"ottingen. The HET is named in honor of its principal benefactors, William P. Hobby and Robert E. Eberly.

\end{document}